\documentclass{sig-alternate}

\usepackage{amsmath}
\usepackage{multirow}
\usepackage{algorithmic}
\usepackage{subfigure}
\usepackage{url}
\usepackage{graphicx}

\makeatletter
\newif\if@restonecol
\makeatother

\usepackage[ruled]{algorithm2e}

\newcommand{\ignore}[1]{}

\begin{document}

\title{Runtime Optimizations for Prediction\\with Tree-Based Models}

\numberofauthors{3}
\author{Nima~Asadi$^{1,2}$, Jimmy~Lin$^{1,2,3}$, Arjen~P.~de~Vries$^4$\\[1ex]
\affaddr{$^1$Dept. of Computer Science, $^{2}$Institute for Advanced Computer Studies,
$^{3}$The iSchool}\\
\affaddr{University of Maryland, College Park}\\[1ex]
\affaddr{$^4$Centrum Wiskunde and Informatica (CWI), Amsterdam}\\[1ex]
\email{nima@cs.umd.edu, jimmylin@umd.edu, arjen@acm.org}
}

\maketitle

\begin{abstract}
Tree-based models have proven to be an effective solution for
web ranking as well as other problems in diverse domains. This
paper focuses on optimizing the runtime performance of applying such models to
make predictions, given an already-trained model. Although exceedingly simple conceptually, most implementations
of tree-based models do not efficiently utilize modern superscalar processor
architectures. By laying out data structures in memory in a more
cache-conscious fashion, removing branches from the execution flow
using a technique called predication, and micro-batching predictions
using a technique called vectorization, we are able to better exploit modern
processor architectures and significantly improve the speed of
tree-based models over hard-coded if-else blocks. Our work contributes to
the exploration of {\it architecture-conscious} runtime
implementations of machine learning algorithms.
\end{abstract}

\section{Introduction}

Recent studies have shown that machine-learned tree-based models, combined with
ensemble techniques, are highly effective for building web ranking
algorithms~\cite{Burges_2010,Ganjisaffar_etal_SIGIR2011,Tyree_etal_WWW2011}
within the ``learning to rank'' framework~\cite{LiHang_2011}.
Beyond document retrieval, tree-based models have also been
proven effective for tackling problems in diverse domains such as
online advertising~\cite{Panda_etal_VLDB2009}, medical
diagnosis~\cite{Ge_Wong_2008}, genomic
analysis~\cite{Schietgat_etal_2010}, and computer
vision~\cite{Criminisi_etal_2011}. This paper focuses on runtime
optimizations of tree-based models that take advantage of modern
processor architectures:\ we assume that a model has already been
trained, and now we wish to make predictions on new data as fast as possible.
Although exceedingly simple, tree-based
models do not efficiently utilize modern processor architectures due to the
prodigious amount of branches and non-local memory references in
standard implementations. By laying out data structures in memory in a
more cache-conscious fashion, removing branches from the execution
flow using a technique called predication, and micro-batching predictions
using a technique called vectorization, we are able to better exploit modern
processor architectures and significantly improve the speed of
tree-based models over hard-coded if-else blocks.

Our experimental results are measured in nanoseconds for individual
trees and microseconds for complete ensembles. A natural starting question is:\ do
such low-level optimizations actually matter? Does shaving microseconds off an
algorithm have substantive impact on a real-world task? We argue that the answer is {\it
  yes}, with two different motivating examples:\ First, in our primary
application of learning to rank for web search, prediction by
tree-based models forms the inner loop of a search engine. Since
commercial search engines receive billions of queries per day,
improving this tight inner loop (executed, perhaps, many billions of
times) can have a noticeable effect on the bottom line. Faster prediction
translates into fewer servers for the same query load, reducing datacenter
footprint, electricity and cooling costs, etc. Second, in the
domain of financial engineering, every nanosecond
counts in high frequency trading.
Orders on NASDAQ are fulfilled in less than 40 microseconds.\footnote{\small http://www.nasdaqtrader.com/Trader.aspx?id=colo}
Firms fight over
the length of cables due to speed-of-light propagation delays,
both within an individual datacenter and across oceans~\cite{Johnson_etal_2012}.\footnote{\small http://spectrum.ieee.org/computing/it/financial-trading-at-the-speed-of-light} 
Thus, for machine learning in financial
engineering, models that shave even a few microseconds off
prediction times present an edge.

We view our work as having the following contributions:\ First, we
introduce the problem of {\it architecture-conscious} implementations
of machine learning algorithms to the information retrieval and data
mining communities. Although similar work has long existed in the database
community~\cite{Ailamaki_etal_VLDB1999,RaoJun_Ross_VLDB1999,RossKenneth_etal_2005,Zukowski_etal_2005,Boncz_etal_CACM2008},
there is little research on the application of
architecture-conscious optimizations for information retrieval and machine learning
problems. Second, we propose novel implementations of tree-based
models that are highly-tuned to modern processor architectures, taking
advantage of cache hierarchies and superscalar processors. Finally, we
illustrate our techniques in a standard, widely-accepted, learning-to-rank task and show
significant performance improvements over standard implementations and
hard-coded if-else blocks.


\section{Background and Related Work}
\label{section:bg}

We begin with an overview of modern processor architectures and recap
advances over the past few decades. The broadest trend is perhaps the
multi-core revolution~\cite{Olukotun_Hammond_2005}:\ the relentless
march of Moore's Law continues to increase the number of transistors
on a chip exponentially, but experts widely agree that we
are long past the point of diminishing returns in extracting
instruction-level parallelism in hardware. Instead, adding more
cores appears to be a better use of increased transistor
density.
Since prediction is an embarrassingly parallel problem, our techniques can
ride the wave of increasing core counts.

A less-discussed, but just as important trend over the past two
decades is the so-called ``memory wall''~\cite{Boncz_etal_CACM2008},
where increases in processor speed have far outpaced improvements in memory
latency. This means that RAM is becoming slower relative to the CPU. In the
1980s, memory latencies were on the order of a few clock cycles;
today, it could be several hundred clock cycles. To hide this latency,
computer architects have introduced hierarchical cache memories:\ a
typical server today will have L1, L2, and L3 caches between the
processor and main memory. Cache architectures are built on the
assumption of reference locality---that at any given time, the
processor repeatedly accesses only a (relatively) small amount of data,
and these fit into cache. The fraction of memory accesses that can be
fulfilled directly from the cache is called the {\it cache hit rate},
and data not found in cache is said to cause a {\it cache miss}. Cache
misses cascade down the hierarchy---if a datum is not found in L1, the
processor tries to look for it in L2, then in L3, and finally in main
memory (paying an increasing latency cost each level down).

Managing cache content is a complex challenge, but there are two main
principles that are relevant to a software developer. First, caches
are organized into cache lines (typically 64 bytes), which is the
smallest unit of transfer between cache levels. That is, when a
program accesses a particular memory location, the entire cache line
is brought into (L1) cache. This means that subsequent references to
nearby memory locations are very fast, i.e., a cache hit. Therefore,
in software it is worthwhile to organize data structures to take
advantage of this fact. Second, if a program accesses memory in a
predictable sequential pattern (called striding), the processor will
prefetch memory blocks and move them into cache, before the program has
explicitly requested the memory locations (and in certain architectures, it is possible
to explicitly control prefetch in software). There is, of course, much more
complexity beyond this short description; see~\cite{Jacob_2009} for an
overview.

The database community has explored in depth the consequences of
modern processor architectures for relational query
processing~\cite{Ailamaki_etal_VLDB1999,RaoJun_Ross_VLDB1999,RossKenneth_etal_2005,Zukowski_etal_2005,Boncz_etal_CACM2008}. In
contrast, these issues are underexplored for information retrieval and
data mining applications. This is one of the first attempts
at developing architectural-conscious runtime implementations of
machine learning algorithms. Researchers have explored
scaling the {\it training} of tree-based models to massive
datasets~\cite{Panda_etal_VLDB2009,Svore_Burges_2011}, which is of
course an important problem, but orthogonal to the issue we tackle
here:\ given a trained model, how do we make predictions quickly?

Another salient property of modern CPUs is pipelining, where
instruction execution is split between several stages (modern
processors have between one to two dozen stages). At each clock
cycle, all instructions ``in flight'' advance one stage in the
pipeline; new instructions enter the pipeline and instructions that
leave the pipeline are ``retired''. Pipeline stages allow faster clock
rates since there is less to do per stage. Modern {\it superscalar}
CPUs add the ability to dispatch multiple instructions per clock
cycle (and out of order) provided that they are independent.

Pipelining suffers from two dangers, known as ``hazards'' in
VLSI design terminology. {\it Data hazards} occur when one instruction
requires the result of another (that is, a data dependency). This
happens frequently when dereferencing pointers, where we must first
compute the memory location to access. Subsequent instructions cannot
proceed until we actually know what memory location we are
accessing---the processor simply stalls waiting for the result (unless there are other
independent instructions that can be executed). {\it
  Control hazards} are instruction dependencies introduced by if-then
clauses (which compile to conditional jumps in assembly).
To cope with this, modern processors use {\it
  branch prediction techniques}---in short, trying to predict which
code path will be taken. However, if the guess is not correct, the
processor must ``undo'' the instructions that occurred after the
branch point (``flushing'' the pipeline).

The impact of data and control hazards
can be substantial:\ an influential paper in 1999
concluded that in commercial RDBMSes at the time, almost half of the
execution time is spent on stalls~\cite{Ailamaki_etal_VLDB1999}.\footnote{\small Of course,
  this was before the community was aware of the issue, and so systems
  have become much more efficient since then.}
Which is ``worse'', data or control hazards? Not surprisingly, the
answer is, it depends. However, with a technique called
predication~\cite{August_etal_1997,KimHyesoon_etal_2006}, which we explore in our work, it is possible to
convert control dependencies into data dependencies (see
Section~\ref{section:approach}). Whether predication is worthwhile,
and under what circumstances, remains an empirical question.

Another optimization that we adopt, called vectorization, was
pioneered by database
researchers~\cite{Boncz_etal_CIDR2005,Zukowski_etal_2005}:\ the basic
idea is that instead of processing a tuple at a time, a relational query
engine should process a ``vector'' (i.e., batch) of tuples at a time to
take advantage of pipelining.\footnote{\small Note that this sense of vectorization
is distinct from, but related to, explicit SIMD instructions that are available in
many processor architectures today. Vectorization increases the opportunities
for optimizing compilers to generate specialized SIMD instructions automatically.} Our work represents the first
application of vectorization to optimizing machine learning algorithms that
we are aware of.

Beyond processor architectures, the other area of relevant
work is the vast literature on learning to rank~\cite{LiHang_2011},
application of machine learning techniques to document ranking in search.
Our work uses gradient-boosted regression trees
(GBRTs)~\cite{Burges_2010,Tyree_etal_WWW2011,Ganjisaffar_etal_SIGIR2011},
a state-of-the-art ensemble method. The focus of most
learning-to-rank research is on learning effective models, without
considering efficiency, although there is an emerging thread of
work that attempts to better balance both factors~\cite{Wang_etal_SIGIR2011,XuZhixiang_etal_ICML2012}.
In contrast, we focus exclusively on runtime ranking
performance, assuming a model that has already been trained (by
other means).

\section{Tree Implementations}
\label{section:approach}

In this section we describe various implementations of tree-based
models, starting from two baselines and progressively introducing
architecture-conscious optimizations. We focus on
an individual tree, the runtime execution of which involves checking a
predicate in an interior node, following the left or right branch
depending on the result of the predicate, and repeating until a leaf
node is reached. We assume that the predicate at each node involves a
feature and a threshold:\ if the feature value is less than the
threshold, the left branch is taken; otherwise, the right branch is
taken. Of course, trees with greater branching factors and more complex
predicate checks can be converted into an equivalent binary tree, so
our formulation is general.
Note that our discussion is agnostic with respect to the
predictor at the leaf node, be it a boolean (in the classification
case), a real (in the regression case), or even an embedded sub-model.

We assume that the input feature vector is densely-packed in a
floating-point array (as opposed to a sparse, map-based
representation). This means that checking the predicate at each tree
node is simply an array access, based on a unique
consecutively-numbered id associated with each feature.

\smallskip \noindent \textsc{Object}:
As a high-flexibility baseline, we consider an implementation of trees
with nodes and associated left and right pointers in
C++. Each tree node is represented by an object, and contains the
feature id to be examined as well as the decision threshold.
For convenience, we refer to this as the \textsc{Object} implementation.
In our mind, this represents the most obvious
implementation of tree-based models that a software engineer would
come up with---and thus serves as a good point of comparison.

This implementation has two advantages:\ simplicity and flexibility.
However, we have no control over the physical layout of the tree nodes in memory,
and hence no guarantee that the data structures exhibit good reference locality.
Prediction with this implementation essentially boils down to pointer chasing
across the heap:\ when following either the left or the right pointer
to the next tree node, the processor is likely to be stalled by a cache miss.

\smallskip \noindent \textsc{CodeGen:}
As a high-performance baseline, we consider statically-generated if-else
blocks. That is, a code generator takes a tree model and directly
generates C code, which is then compiled and used to make predictions.
For convenience, this is referred to as the \textsc{CodeGen}
implementation.
This represents the most obvious performance optimization that
a software engineer would come up with, and thus serves as another
good point for performance comparison.

We expect this approach to be fast. The entire model is statically
specified; machines instructions are expected to be relatively compact
and will fit into the instruction cache, thus exhibiting
good reference locality. Furthermore, we leverage decades of compiler
optimizations that have been built into GCC. Note that this eliminates
data dependencies completely by converting them all into control
dependencies.

The downside, however, is that this approach is inflexible.
The development cycle now requires more steps:\ after training
the model, we need to run the code generation, compile the resulting
code, and then link against the rest of the system. This may be a
worthwhile tradeoff for a production system,
but from the view of rapid experimentation and iteration, the
approach is a bit awkward.

\smallskip \noindent \textsc{Struct:} The \textsc{Object} approach has
two downsides:\ poor memory layout (i.e., no reference locality and
hence cache misses) and inefficient memory utilization (due to
object overhead). To address the second point, the solution is
fairly obvious:\ get rid of C++ and drop down to C to avoid the object
overhead. We can implement each node as a {\tt \small struct} in C
(comprising feature id, threshold, left and right pointers). We
construct a tree by allocating memory for each node ({\tt \small
  malloc}) and assigning the pointers appropriately. Prediction with
this implementation remains an exercise in pointer chasing, but now across
more memory-efficient data structures. We refer to this as the
\textsc{Struct} implementation.

\smallskip \noindent \textsc{Struct$^+$}: An improvement over the
\textsc{Struct} implementation is to physically manage the memory
layout ourselves. Instead of allocating memory for each node
individually, we allocate memory for all the nodes at once (i.e., an
array of {\tt \small struct}s) and linearize the tree in the following
way:\ the root lies at index 0. Assuming a perfectly-balanced tree, for a node at index $i$, its left
child is at $2i + 1$ and its right child is at $2i + 2$. 
This is equivalent to laying out the tree using a breadth-first traversal of the nodes.
The hope is that by manually controlling memory layout, we can achieve better
reference locality, thereby speeding up the memory
references. This is similar to the idea behind CSS-Trees~\cite{RaoJun_Ross_VLDB1999}
used in the database community.
For convenience we call this the \textsc{Struct$^+$} implementation.

One nice property of retaining the left and right pointers in this
implementation is that for unbalanced trees (i.e., trees with missing
nodes), we can more tightly pack the nodes to remove ``empty
space'' (still following the layout approach based on breadth-first node traversal).
Thus, the \textsc{Struct$^+$} implementation occupies the
same amount of memory as \textsc{Struct}, except that the memory is
contiguous.

\smallskip \noindent \textsc{Pred:}
The \textsc{Struct$^+$} implementation tackles the reference
locality problem, but there remains one more issue:\
the presence of branches (resulting from the conditionals), which
can be quite expensive to execute. Branch
mispredicts may cause pipeline stalls and wasted cycles (and of
course, we would expect many mispredicts with trees). Although it is true that
speculative execution renders the situation far more
complex, removing branches may yield performance increases.
A well-known trick in the compiler community for overcoming these issues is known
as predication~\cite{August_etal_1997,KimHyesoon_etal_2006}.
The underlying idea is to convert
control dependencies (hazards) into data dependencies (hazards), thus altogether
avoiding jumps in the underlying assembly code.

Here is how predication is adapted for our case: We encode the tree as
a {\tt \small struct} array in C, {\tt \small nd}, where {\tt \small
  nd[i].fid} is the feature id to examine, and {\tt \small
  nd[i].theta} is the threshold.
We assume a
fully-branching binary tree, with nodes laid out via breadth-first
traversal (i.e., for a node at index $i$, its left
child is at $2i + 1$ and its right child is at $2i + 2$).
To make the prediction, we probe the array in the following manner:
\begin{quote}{\scriptsize
\smallskip
\begin{verbatim}
i = (i<<1) + 1 + (v[nd[i].fid] >= nd[i].theta);
i = (i<<1) + 1 + (v[nd[i].fid] >= nd[i].theta);
  ...
\end{verbatim}}
\end{quote}
\noindent We completely unroll the tree traversal loop, so the
above statement is repeated $d$ times for a tree of depth $d$. At the
end, $i$ contains the index of the leaf node corresponding to the
prediction (which we look up in another array).
One final implementation
detail:\ we hard code a prediction function for each tree depth, and then
dispatch dynamically using function pointers.
Note that this approach assumes a
fully-balanced binary tree; to cope with unbalanced trees, we expand
by inserting dummy nodes.

\smallskip \noindent \textsc{VPred:}
Predication eliminates branches but at the cost of introducing
data hazards. Each statement in \textsc{Pred} requires an indirect
memory reference. Subsequent instructions cannot execute until the
contents of the memory locations are fetched---in other words, the processor
will simply stall waiting for memory references to resolve. Therefore, predication is
entirely bottlenecked on memory access latencies.

A common technique adopted in the database literature to mask these
memory latencies is called {\it vectorization}~\cite{Boncz_etal_CIDR2005,Zukowski_etal_2005}. Applied to our task,
this translates into operating on multiple instances (feature vectors) at once, in an
interleaved way. This takes advantage of multiple dispatch and pipelining in
modern processors (provided that there are no dependencies between dispatched
instructions, which is true in our case). So, while the processor is waiting for the
memory access from the predication step on the first instance, it can
start working on the second instance. In fact, we can work on $v$
instances in parallel. For $v=4$, this looks like the following,
working on instances {\tt \small i0}, {\tt \small i1}, {\tt \small
  i2}, {\tt \small i3} in parallel:
\begin{quote}{\scriptsize
\begin{verbatim}
i0 = (i0<<1) + 1 + (v[nd[i0].fid] >= nd[i0].theta);
i1 = (i1<<1) + 1 + (v[nd[i1].fid] >= nd[i1].theta);
i2 = (i2<<1) + 1 + (v[nd[i2].fid] >= nd[i2].theta);
i3 = (i3<<1) + 1 + (v[nd[i3].fid] >= nd[i3].theta);

i0 = (i0<<1) + 1 + (v[nd[i0].fid] >= nd[i0].theta);
i1 = (i1<<1) + 1 + (v[nd[i1].fid] >= nd[i1].theta);
i2 = (i2<<1) + 1 + (v[nd[i2].fid] >= nd[i2].theta);
i3 = (i3<<1) + 1 + (v[nd[i3].fid] >= nd[i3].theta);
  ...
\end{verbatim}}
\end{quote}
\noindent In other words, we traverse one layer in the tree for four
instances at once. While we're waiting for {\tt \small v[nd[i0].fid]}
to resolve, we dispatch instructions for accessing {\tt \small
  v[nd[i1].fid]}, and so on. Hopefully, by the time the final memory
access has been dispatched, the contents of the first memory access
are available, and we can continue without processor stalls.

Again, we completely unroll the tree traversal loop, so each
block of statements is repeated $d$ times for a tree of depth $d$. At the
end, $i$ contains the index of the leaf nodes corresponding to the
prediction for $v$ instances. Setting $v$ to 1 reduces this model to
pure predication (i.e., no vectorization). Note that the optimal value
of $v$ is dependent on the relationship between the amount of computation
performed and memory latencies---we will determine this relationship empirically.
For convenience, we refer to the vectorized version of the predication
technique as \textsc{VPred}.

\section{Experimental Setup}
\label{section:experimental_setup}

Given that the focus of our work is efficiency, our primary
evaluation metric is prediction speed. We define this as the
elapsed time between the moment a feature vector (i.e., a test instance)
is presented to the tree-based model to the moment that a prediction
(in our case, a regression value) is made for the
instance. To increase the reliability of our results, we conducted multiple trials
and report the mean and variance.

We conducted two sets of experiments:\ first, using syn\-thetically-generated
data to quantify the performance of individual
trees in isolation, and second, on standard learning-to-rank datasets to
verify the performance of full ensembles.

All experiments were run on a Red Hat Linux server, with
Intel Xeon Westmere quad-core processors (E5620 2.4GHz).
This architecture has a 64KB L1 cache per core,
split between data and instructions; a 256KB L2 cache per core; and a
12MB L3 cache shared by all cores. Code was
compiled with GCC (version 4.1.2) using optimization flags
{\tt \small -O3 -fomit-frame-pointer -pipe}. All code ran single-threaded.

\subsection{Synthetic Data}
\label{section:experimental_setup:synthetic}

The synthetic data consisted of
randomly generated trees and randomly generated feature vectors.
Each intermediate node in a tree has two fields:\ a feature id and a threshold
on which the decision is made. Each leaf is associated with a regression value.
Construction of a random tree of depth $d$ begins with the root node.
We pick a feature id at random and generate a random
threshold to split the tree into left and right subtrees.
This process is recursively performed to build each subtree
until we reach the desired tree depth. When we reach a leaf node,
we generate a regression value at random. Note that our
randomly-generated trees are fully-balanced, i.e., a tree of depth $d$ has $2^d$ leaf nodes.

Once a tree has been constructed, the next step is to generate random feature vectors.
Each random feature vector is simply a floating-point array of length $f$ ($=$ number of features),
where each index position corresponds to a feature value.
We assume that all paths in the decision tree are equally likely;
the feature vectors are generated in a way that guarantees
an equal likelihood of visiting each leaf.
To accomplish this, we take one leaf at a time
and follow its parents back to the root. At each node, we take the node's
feature id and produce a feature value based on the position of the child
node. That is, if the child node we have just visited is on the left subtree
we generate a feature value that is smaller than the threshold stored at the
current parent node; otherwise we generate a feature value larger than the threshold.
We randomize the order of instances once we have generated all the feature vectors. 
To avoid any cache effects, our experiments are conducted on a large number of instances (512k).

Given a random tree and a set of random feature vectors, we ran
experiments to assess the various implementations of tree-based models
described in Section~\ref{section:approach}.
To get a better sense of the variance, we performed 5 trials;
in each trial we constructed a new random binary tree and a different
randomly-generated set of feature vectors. To explore the design space,
we conducted experiments with varying tree depths $d \in \{3, 5, 7, 9, 11\}$
and varying feature sizes $f \in \{32, 128, 512\}$.

\begin{figure*}[t]
\centering
\subfigure[$f=32$]{\label{figure:results:synthetic:32}\includegraphics[width=0.32\linewidth]{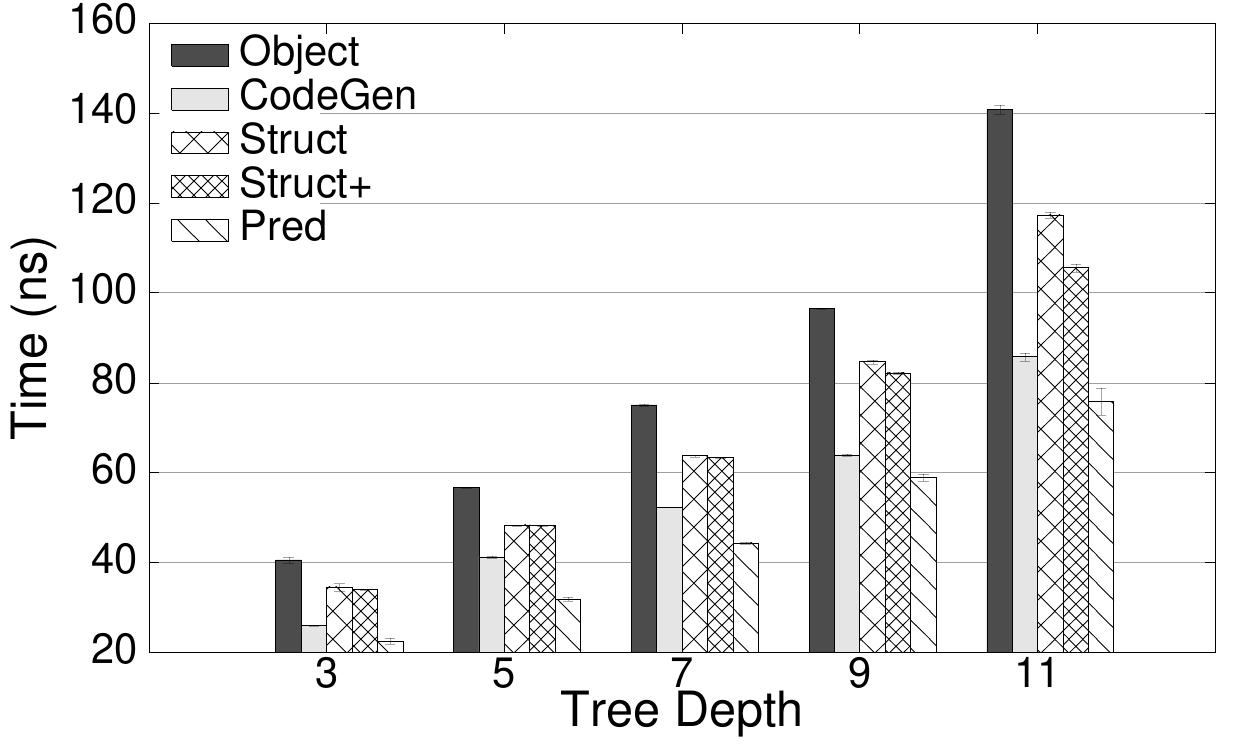}}
\subfigure[$f=128$]{\label{figure:results:synthetic:128}\includegraphics[width=0.32\linewidth]{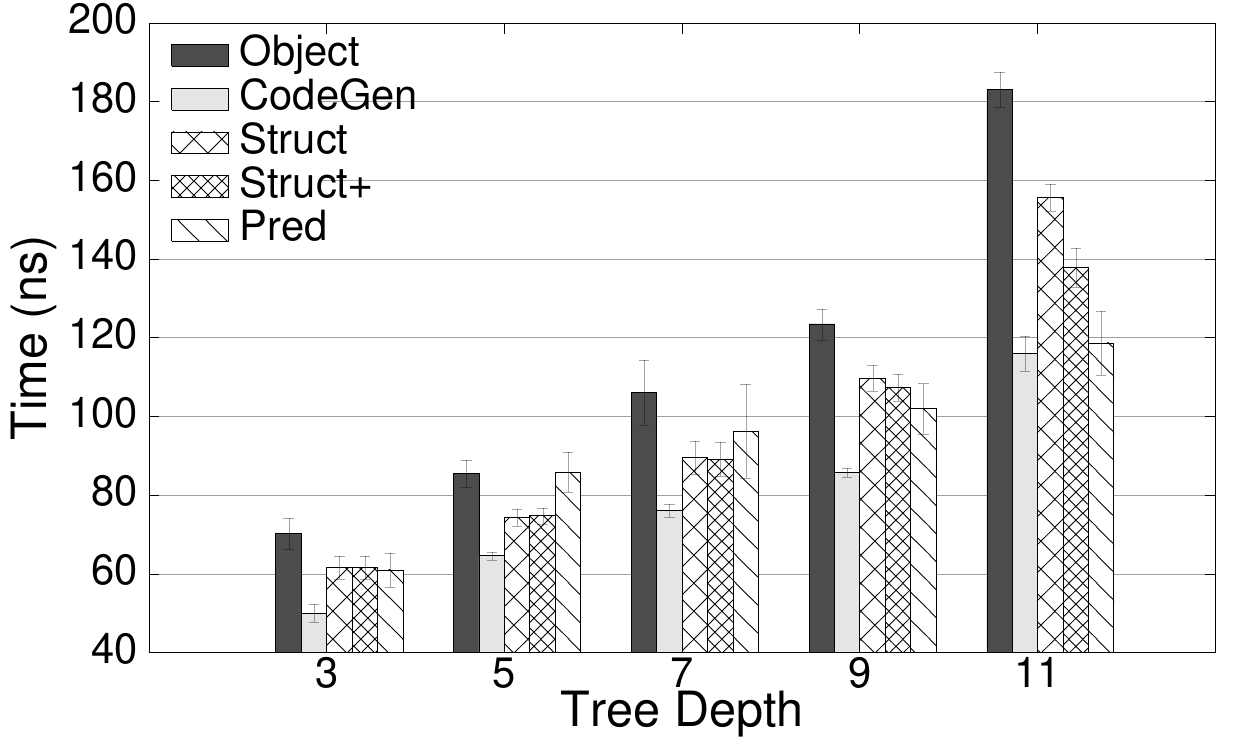}}
\subfigure[$f=512$]{\label{figure:results:synthetic:512}\includegraphics[width=0.32\linewidth]{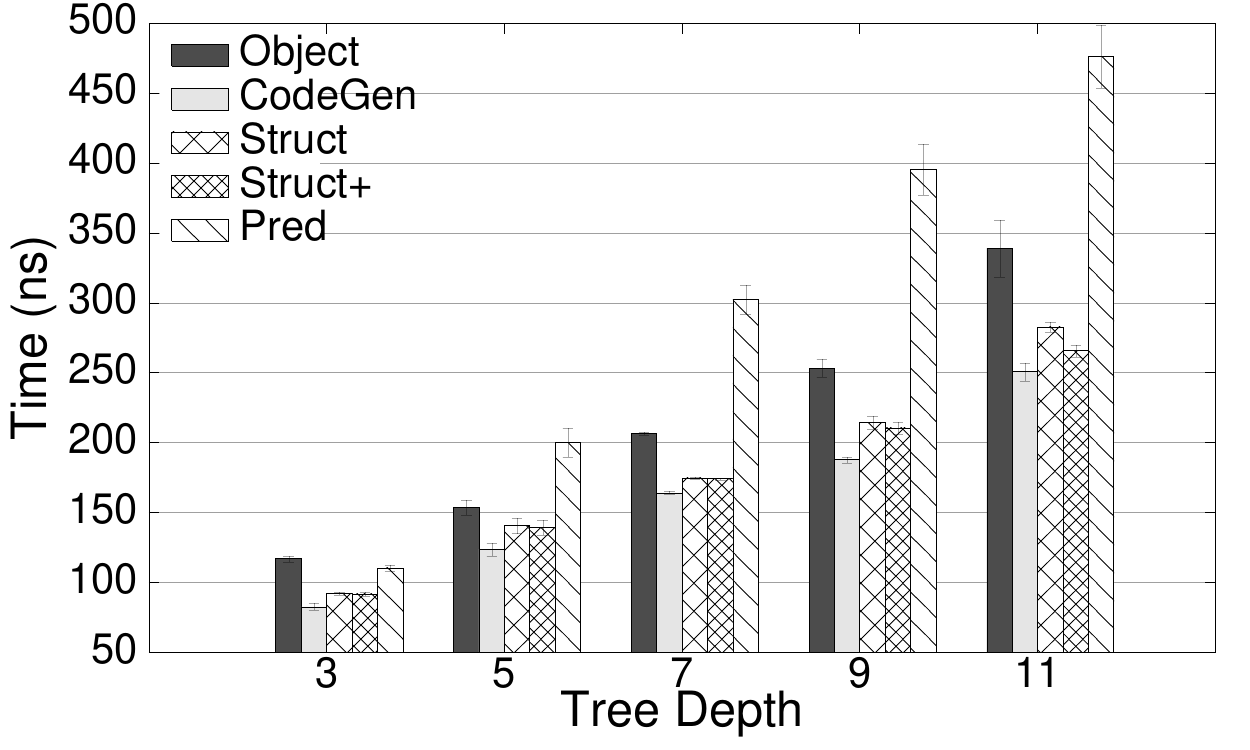}}
\caption{Prediction time per instance (in nanoseconds) on synthetic data using various implementations.}
\label{figure:results:synthetic}
\end{figure*}

\subsection{Learning-to-Rank Experiments}

In addition to randomly-generated trees, we conducted experiments using
standard learning-to-rank datasets, where training, validation, and test
data are provided. Using the
training and validation sets we learned a complete tree-ensemble ranking model,
and evaluation is then carried out on test instances
to determine the speed of the various implementations.
These experiments assess performance in a real-world application.

We used gradient-boosted regression trees (GBRTs)~\cite{Burges_2010,Tyree_etal_WWW2011,Ganjisaffar_etal_SIGIR2011}
to train a learning-to-rank model. GBRTs are ensembles of regression trees
that yield state-of-the-art effectiveness on learning-to-rank tasks.
The learning algorithm sequentially adds new trees to the ensemble
that best account for the remaining regression error (i.e., the residuals).
We used the open-source jforests 
implementation\footnote{\small http://code.google.com/p/jforests/}
of LambdaMART to optimize NDCG~\cite{Jarvelin_Kekalainen_TOIS2002}. Although there is no
way to precisely control the depth of each tree,
we can adjust the size distribution of the trees by
setting a cap on the number of leaves (which is an input parameter to the learner).

We used two standard learning-to-rank datasets:
LETOR-MQ2007\footnote{\small
  http://research.microsoft.com/en-us/um/beijing/projects/letor/letor4dataset.aspx}
and MSLR-WEB10K.\footnote{\small
  http://research.microsoft.com/en-us/projects/mslr/}
Both are pre-folded, providing training, validation, and test
instances.  Table~\ref{table:datasets} shows the dataset sizes and the
numbers of features. To measure variance, we repeated experiments on
all five folds. Note that MQ2007 is much smaller 
and is considered by many in the community to
be outdated.

\begin{table}[t]
\caption{Average number of training, validation, and test instances
in our learning-to-rank datasets, along with the number of features.
\label{table:datasets}}
\begin{center}
\begin{tabular}{|l|r|r|r||r|}
\hline
Dataset & Train & Validate & Test & Features\\
\hline
\hline
MSLR-WEB10K & 720K & 240K & 240K & 136 \\
LETOR-MQ2007 & 42K & 14K & 14K & 46 \\
\hline
\end{tabular}
\end{center}
\vspace{-0.5cm}
\end{table}

The values of $f$ (number of features) in our synthetic experiments
are guided by these learning-to-rank datasets.
We selected feature sizes that are multiples
of 16 (4-byte floats) so that the feature vectors are integer multiples of cache line
sizes (64 bytes):\ $f=32$ roughly
corresponds to LETOR features and is representative of a small feature
space; $f=128$ corresponds to MSLR and is representative of a
medium-sized feature space. We introduced a third condition $f=512$ to
capture a large feature space condition.

\section{Results}
\label{section:results}

In this section we present experimental results, beginning with
evaluation on synthetic data and then on learning-to-rank
datasets.

\begin{figure*}[t]
\centering
\subfigure[$f=32$]{\label{figure:results:vectorization:32}\includegraphics[width=0.32\linewidth]{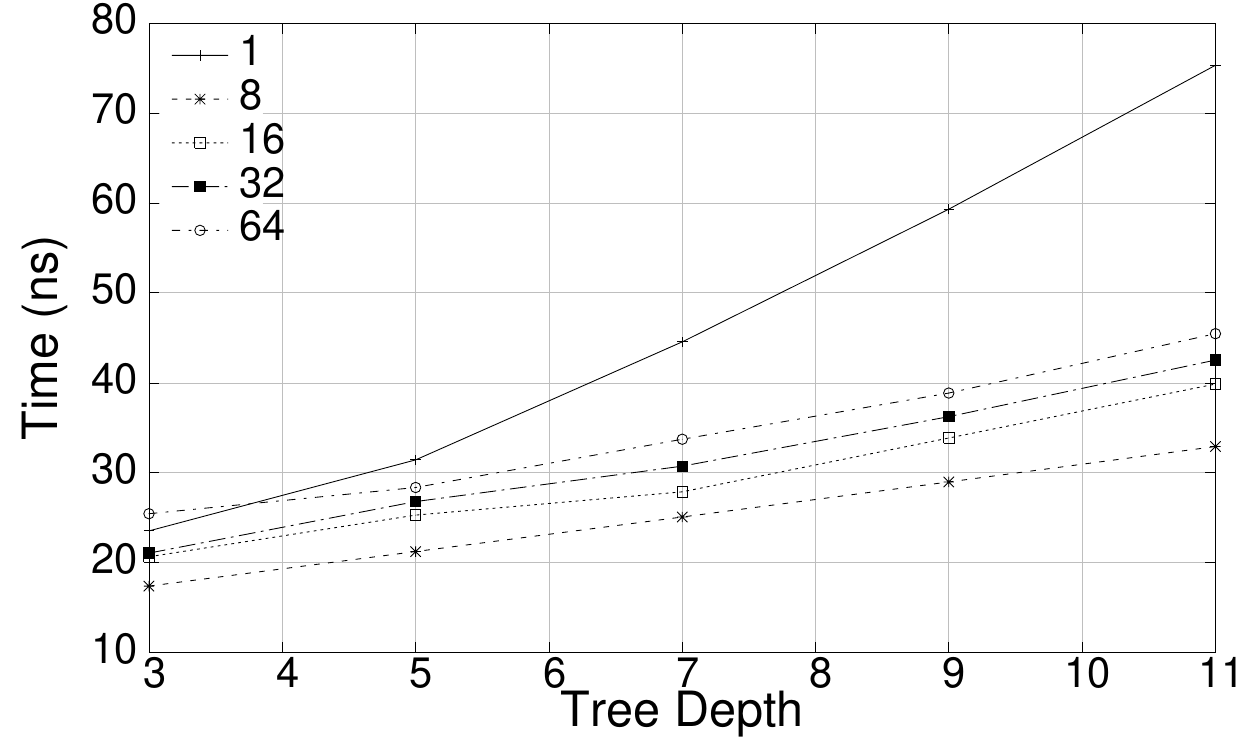}}
\subfigure[$f=128$]{\label{figure:results:vectorization:128}\includegraphics[width=0.32\linewidth]{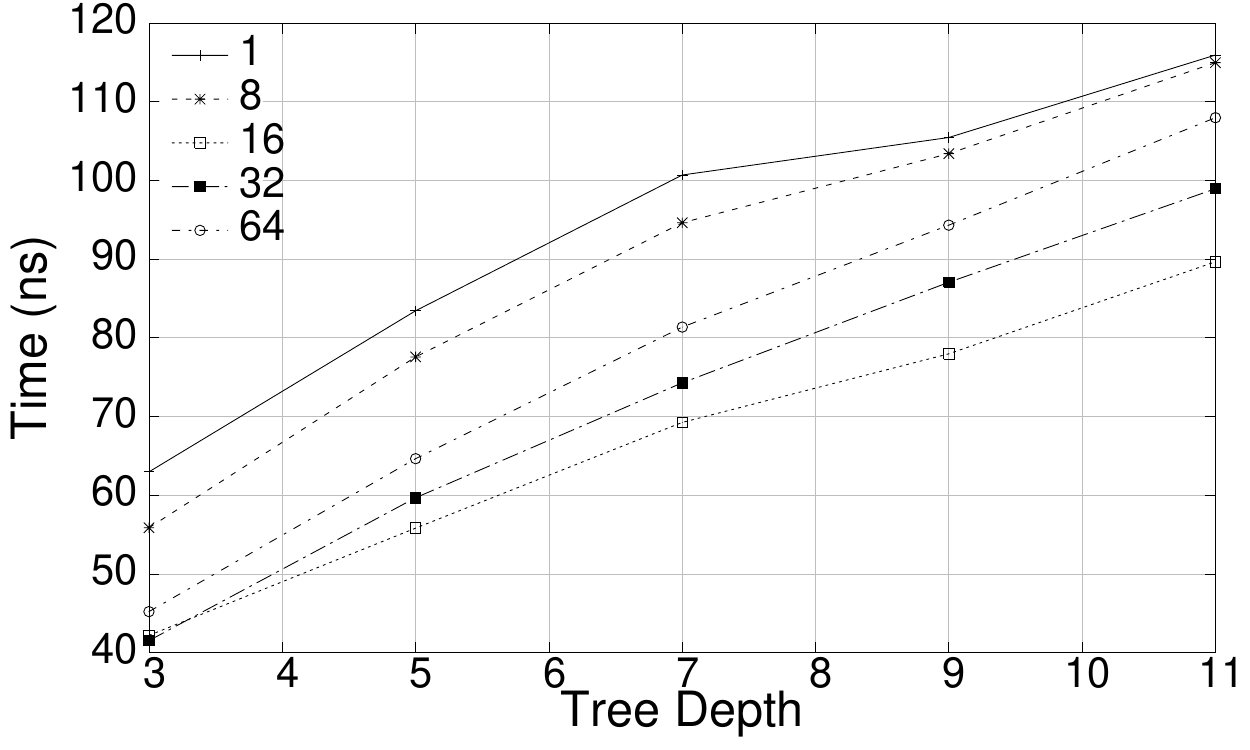}}
\subfigure[$f=512$]{\label{figure:results:vectorization:512}\includegraphics[width=0.32\linewidth]{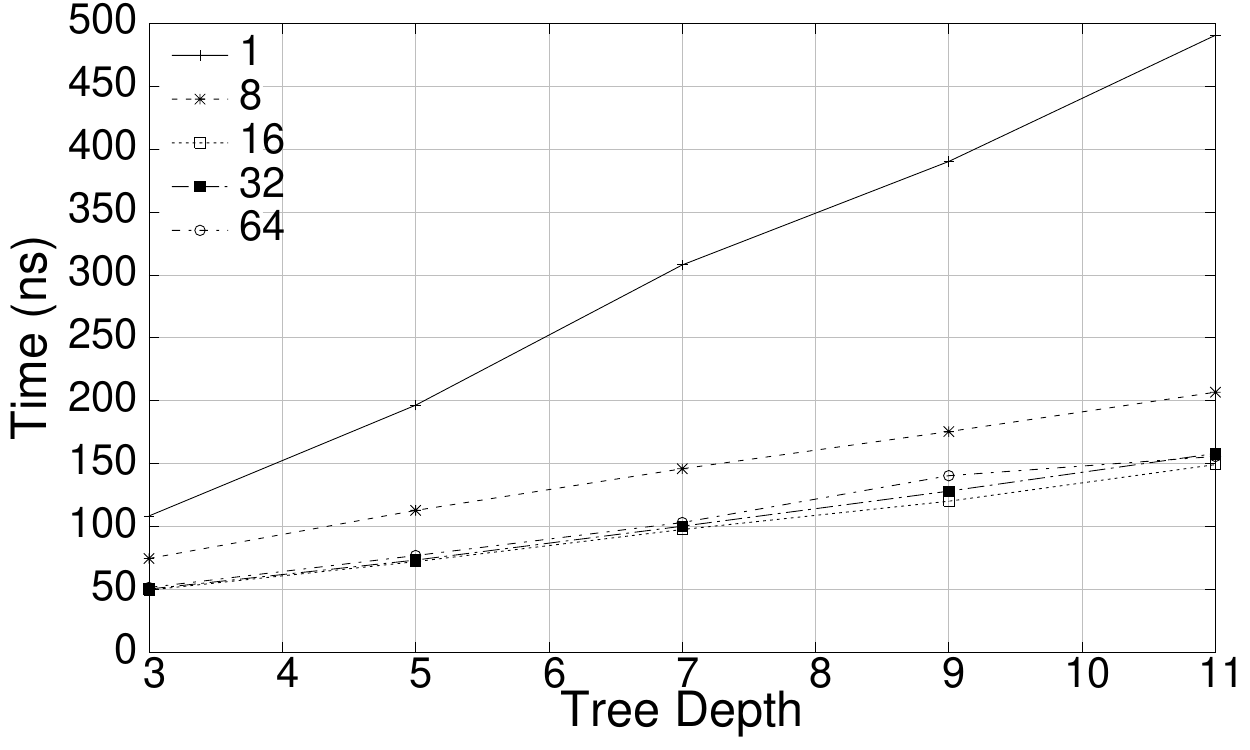}}
\caption{Prediction time per instance (in nanoseconds) on synthetic
  data using vectorized predication, for varying values of the batch size $v$.}
\label{figure:results:vectorization}
\end{figure*}

\subsection{Synthetic Data: Base Results}

We begin by focusing on the first five implementations described in
Section~\ref{section:approach} (leaving aside \textsc{VPred} for now),
using the procedure described in
Section~\ref{section:experimental_setup:synthetic}.
The prediction time per randomly-generated test instance is shown in
Figure~\ref{figure:results:synthetic}, measured in nanoseconds.
The balanced randomly-generated trees vary in terms of tree depth
{\it d}, and each bar chart shows a separate value of {\it f} (number of features).
Time is averaged across five
trials and error bars denote 95\% confidence intervals.
It is clear that as trees become deeper, prediction speeds decrease overall.
This is obvious since deeper trees require more feature accesses and predicate checks, 
more pointer chasing, and more branching (depending on the implementation).

First, consider the high-flexibility and high-performance baselines.
As expected, the \textsc{Object} implementation is the slowest (except for \textsc{Pred} with $f=512$).
It is no surprise that the C++ implementation is slow 
due to the overhead from classes and objects (recall the other implementations are in C).
The gap between \textsc{Object} and \textsc{Struct}, which is the comparable
C implementation, grows with larger trees.
Also as expected, the \textsc{CodeGen} implementation is very fast:\ with
the exception of $f=32$, hard-coded if-else statements are faster or just as fast as all other
implementations, regardless of tree depth.

Comparing \textsc{Struct$^+$} with \textsc{Struct}, we observe no
significant improvement for shallow trees, but a significant speedup
for deep trees. Recall that in \textsc{Struct$^+$}, we
allocate memory for the entire tree so that it resides in a
contiguous memory block, whereas in \textsc{Struct} we let {\tt \small \mbox{malloc}}
allocate memory however it chooses. This shows that reference
locality is important for deeper trees.

Finally, turning to the \textsc{Pred} condition, we observe a very
interesting behavior. For small feature vectors $f=32$, the technique
is actually faster than \textsc{CodeGen}. This shows that
for small feature sizes, predication helps to overcome branch
mispredicts, i.e., converting control dependencies into data dependencies
increases performance. For $f=128$, results are mixed compared to
\textsc{CodeGen}, \textsc{Struct}, and \textsc{Struct$^+$}:\ sometimes
faster, sometimes slower. However, for large feature vectors ($f=512$), the
performance of \textsc{Pred} is terrible, even worse than the
\textsc{Object} implementation. We explain this result as
follows:\ \textsc{Pred} performance is entirely dependent on memory
latency. When traversing the tree, it needs to wait for the
contents of memory before proceeding. Until the memory references are
resolved, the processor simply stalls. With small feature vectors, we
get excellent locality:\ 32 features take up two 64-byte cache lines,
which means that evaluation incurs at most two cache misses. Since
memory is fetched by cache lines, once a feature is accessed, accesses
to all other features on the same cache line are essentially
``free''. Locality decreases as the feature vector size
increases:\ the probability that the predicate at a tree node accesses
a feature close to one that has already been accessed goes down.
Thus, as the feature vector size grows, the \textsc{Pred} prediction time
becomes increasingly dominated by stalls waiting for memory fetches.

The effect of this ``memory wall'' is evident in the other
implementations as well. We observe that the performance differences
between \textsc{CodeGen}, \textsc{Struct}, and \textsc{Struct$^+$}
shrink as the feature size increases (whereas they are more
pronounced for smaller feature vectors). This is because as feature vector
size increases, more and more of the prediction time is dominated by
memory latencies.

How can we overcome these memory latencies? Instead of simply stalling
while we wait for memory references to resolve, we can try to do other
useful computation---this is exactly what vectorization is designed to
accomplish.

\subsection{Tuning Vectorization Parameter}

In Section~\ref{section:approach}, we proposed {\it vectorization}
of the predication technique in order to mask memory latencies. The idea
is to work on $v$ instances (feature vectors) at the same time,
so that while the processor is waiting for memory access for one instance,
useful computation can happen on another. This takes advantage of pipelining
and multiple dispatch in modern superscalar processors.

The effectiveness of vectorization depends on
the relationship between time spent in actual computation and memory
latencies. For example, if memory fetches take only one
clock cycle, then vectorization cannot possibly help. The longer
the memory latencies, the more we would expect vectorization (larger batch sizes) to help.
However, beyond a certain point, once memory
latencies are effectively masked by vectorization, we would expect
larger values of $v$ to have little impact. In fact, values that are too large
start to bottleneck on memory bandwidth and cache size.

\begin{table}[t]
\caption{Prediction time per instance (in nanoseconds) for the
  vectorized predication implementation, compared to simple
  predication and code generation, along with relative improvements.
\label{table:relative_speed}}

\begin{center}
\begin{small}
\subtable[$f=32, v=8$]{
\begin{tabular}{|l|r||r|r||r|r|}
\hline
$d$ & \textsc{VPred} & \textsc{Pred} & $\Delta$ & \textsc{CodeGen} &  $\Delta$ \\
\hline
\hline
3 & 17.4 & 22.6  & 23\% & 26.0 & 33\% \\
5 & 21.3 & 31.9  & 33\% & 41.3 & 48\% \\
7 & 25.1 & 44.4  & 44\% & 52.4 & 52\% \\
9 & 28.9 & 58.9  & 51\% & 63.9 & 55\% \\
11 & 39.2 & 75.8 & 57\% & 85.8 & 54\% \\
\hline
\end{tabular}
\label{table:relative_speed:32}}
\subtable[$f=128, v=16$]{
\begin{tabular}{|l|r||r|r||r|r|}
\hline
$d$ & \textsc{VPred} & \textsc{Pred} & $\Delta$ & \textsc{CodeGen} & $\Delta$ \\
\hline
\hline
3 & 42.2 & 61.0 & 31\% & 50.0 & 16\% \\
5 & 55.8 & 85.9 & 35\% & 64.6 & 14\% \\
7 & 69.3 & 96.3 & 28\% & 76.2 & 9\% \\
9 & 77.9 & 102.0 & 24\% & 85.8 & 9\% \\
11 & 89.6 & 118.7 & 25\% & 116.0 & 23\% \\
\hline
\end{tabular}
\label{table:relative_speed:128}}
\subtable[$f=512, v=16$]{
\begin{tabular}{|l|r||r|r||r|r|}
\hline
$d$ & \textsc{VPred} & \textsc{Pred} & $\Delta$ & \textsc{CodeGen} & $\Delta$ \\
\hline
\hline
3 & 49.7 & 110.6 & 55\% & 82.8 & 40\% \\
5 & 72.3 & 200.3 & 64\% & 123.9 & 42\% \\
7 & 97.8 & 302.5 & 68\% & 164.2 & 40\% \\
9 & 120.4 & 395.5 & 70\% & 187.8 & 36\% \\
11 & 149.5 & 476.1 & 69\% & 250.7 & 40\% \\
\hline
\end{tabular}
\label{table:relative_speed:512}}
\end{small}
\end{center}
\vspace{-0.5cm}
\end{table}

In Figure~\ref{figure:results:vectorization}, we show the impact of
various batch sizes, $v\in\{1, 8, 16, 32, 64\}$, for the different
feature sizes. Note that when $v$ is set to 1, we evaluate
one instance at a time, which reduces to the \textsc{Pred} implementation.
Prediction speed is measured in nanoseconds and normalized
by batch size (i.e., divided by $v$), so we report {\it per-instance}
prediction time.
For $f=32$, $v=8$ yields the best performance; for
$f=128$, $v=16$ yields the best performance; for $f=512$,
$v=\{16,32,64\}$ all provide approximately the same level of performance.
These results are exactly what we would expect:\ since memory
latencies increase with larger feature sizes, a larger batch size is
needed to mask the latencies.

With the combination of vectorization and predication, \textsc{VPred}
becomes the fastest of all our implementations on the synthetic data.
Comparing Figures~\ref{figure:results:synthetic} and
\ref{figure:results:vectorization}, we see that \textsc{VPred}
(with optimal vectorization parameter) is actually faster than
\textsc{CodeGen}. Table~\ref{table:relative_speed} summarizes this comparison.
Vectorization is up to 70\% faster than the non-vectorized implementation;
\textsc{VPred} can be twice as fast as \textsc{CodeGen}.
In other words, we retain the best of both worlds:\ speed and flexibility,
since the \textsc{VPred} implementation does not require code recompilation.

\subsection{Learning-to-Rank Experiments}

\begin{figure*}[t]
\centering
\subfigure[LETOR-MQ2007]{\label{figure:results:actual:letor}\includegraphics[width=0.49\linewidth]{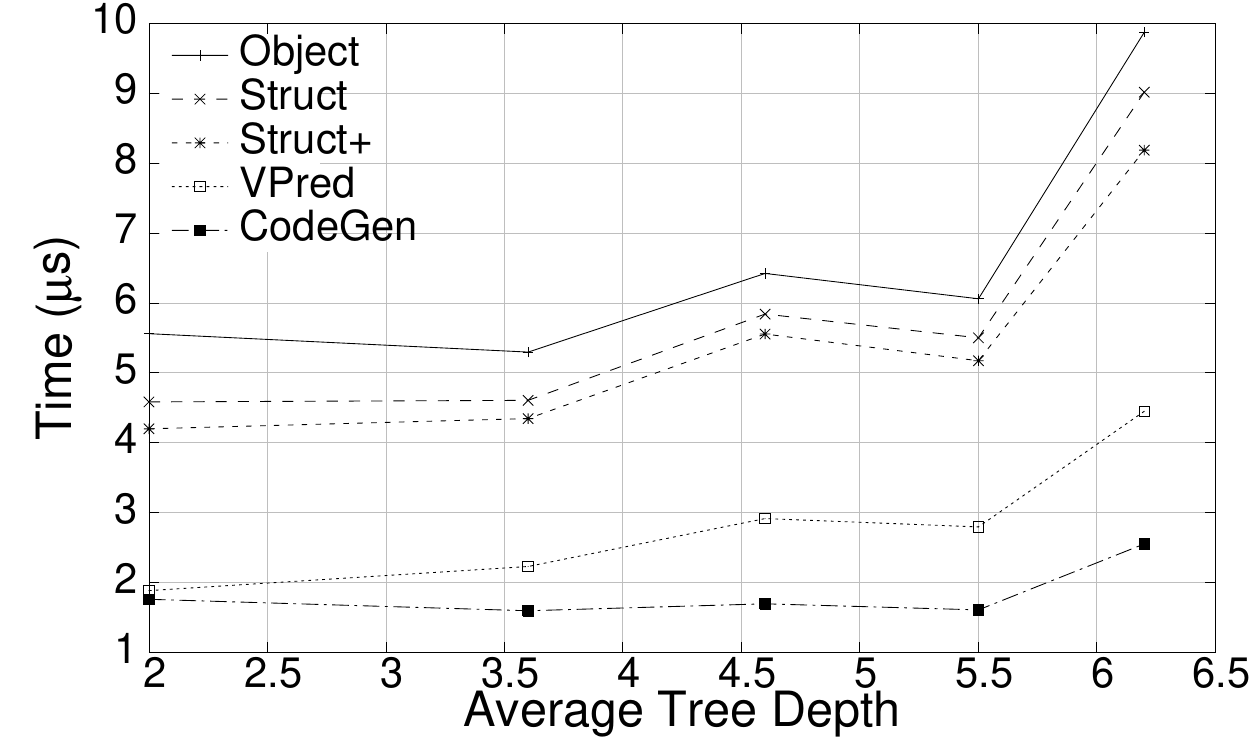}}
\subfigure[MSLR-WEB10K]{\label{figure:results:actual:mslr}\includegraphics[width=0.49\linewidth]{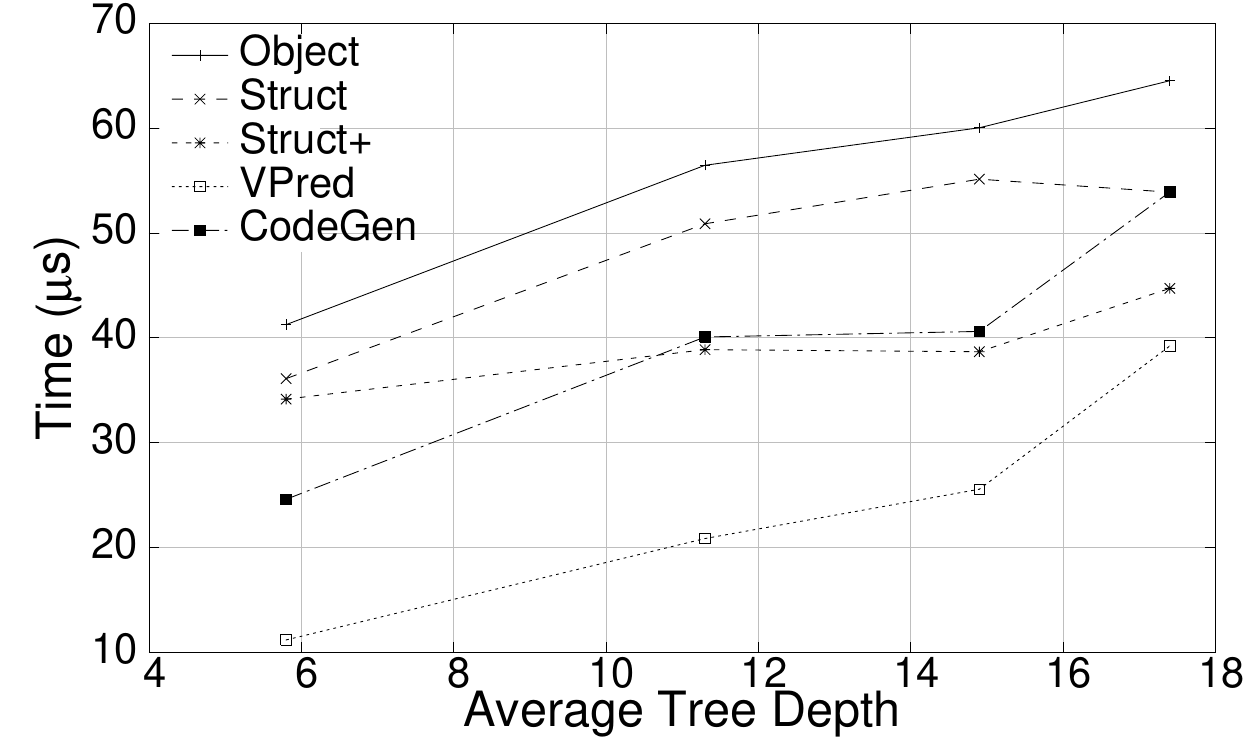}}
\caption{Per-instance prediction times (in microseconds), for
  ensembles of trees trained using LambdaMART on different datasets.}
\label{figure:results:actual}
\end{figure*}

Having evaluated different implementations on synthetic data,
we move on to learning-to-rank datasets using tree ensembles.
As previously described, we used the implementation of LambdaMART by
Ganjisaffar et al.~\cite{Ganjisaffar_etal_SIGIR2011}.
Once a model has been trained and validated, we evaluate on the test set
to measure prediction speed. Since the datasets come pre-folded five
ways, we repeated our experiments five times and report mean and variance
across the runs.

To handle ensembles in our
implementations, we simply add an outer loop to the algorithm that
iterates over individual trees in the ensemble. Note that
Ganjisaffar et al.\ actually construct multiple ensembles, each built
using a random bootstrap of the training data (i.e., {\it bagging}
multiple boosted ensembles). In this work, we do not adopt this
procedure because bagging is embarrassingly parallel from
the runtime execution perspective and hence not particularly
interesting. For learning parameters, we used values
recommended by Ganjisaffar et al.,
with the exception of max leaves (see below). Feature and data
sub-sampling parameters were set to 0.3, minimum percentage of observations per leaf
was set to 0.5, and the learning rate was set to 0.05.

\begin{table}[t]
\caption{NDCG and average tree depth (variance in parentheses) measured across five folds
using various settings for max number of leaves. For MSLR, $^+$ and $^*$ show statistically significant improvements over
models obtained by setting ``Max.\ Leaves'' to 10 and 30 respectively.
\label{table:effectiveness}}
\begin{center}
\begin{small}
\subtable[LETOR-MQ2007]{
\begin{tabular}{|r|r|l|l|l|}
\hline
{\small Max.\ Leaves} & {\small Avg.\ Depth} & @1 & @3 & @20 \\
\hline
\hline
3 & 2.0 (0.0) & 0.469 & 0.476 & 0.590 \\
5 & 3.6 (0.2) & 0.463 & 0.478 & 0.588 \\
7 & 4.6 (0.6) & 0.490 & 0.484 & 0.592 \\
9 & 5.5 (1.0) & 0.475 & 0.477 & 0.589 \\
11 & 6.2 (1.2) & 0.478 & 0.481 & 0.591 \\
\hline
\end{tabular}
\label{table:effectiveness:letor}}
\subtable[MSLR-WEB10K] {
\begin{tabular}{|r|r|l|l|l|}
\hline
{\small Max.\ Leaves} & {\small Avg.\ Depth} & @1 & @3 & @20 \\
\hline
\hline
10 & 5.8 (1.1) & 0.466 & 0.452 & 0.505 \\
30 & 11.3 (6.2) & 0.470$^+$ & 0.456$^+$ & 0.510$^+$ \\
50 & 14.9 (10.7) & 0.475$^+$ & 0.459$^{+}$ & 0.512$^{+*}$ \\
70 & 17.4 (12.9) & 0.466 & 0.453 & 0.510$^{+}$ \\
\hline
\end{tabular}
\label{table:effectiveness:mslr}}
\end{small}
\end{center}
\vspace{-0.5cm}
\end{table}

In terms of performance, shallower trees are naturally preferred. But
what is the relationship between tree depth and ranking effectiveness?
Tree depth with our particular training algorithm cannot be precisely controlled,
but can be indirectly influenced by the maximum number of leaves on an
individual tree (an input to the learner).
Table~\ref{table:effectiveness} shows the average NDCG values (at different ranks)
measured across five folds on the LETOR and MSLR datasets with different values
of this parameter, similar to the range of values explored in~\cite{Ganjisaffar_etal_SIGIR2011}.
Statistical significance was tested using the Wilcoxon test ($p$-value 0.05); none
of the differences on the LETOR dataset were significant. For each condition,
we also report the average depth of the trees that were actually learned.
The average tree depth is computed for every ensemble and
then averaged across the five folds; variance is presented in parentheses.

Results show that for LETOR, tree depth makes no significant difference on NDCG,
whereas larger trees yield better results on MSLR; however, there appears
to be little difference between 50 and 70 max leaves. The results make sense:\
to exploit larger feature spaces we need trees with more nodes. Since many
in the community consider the LETOR dataset to be out of date with an impoverished
feature set, more credence should be given to the MSLR results.

Turning to performance results, Figure~\ref{figure:results:actual} illustrates
per-instance prediction speed for various
implementations on the learning-to-rank datasets. Note that this is on
the entire ensemble, with latencies now measured in microseconds instead of
nanoseconds.
As described above, the trees were trained with different settings of max leaves;
the {\it x}-axis plots the tree depths from Table~\ref{table:effectiveness}.
In this set of experiments, we made use of the \textsc{VPred}
approach with the vectorization parameter set to $8$ for LETOR and $16$ for MSLR.

Results from the synthetic datasets mostly carry over to these
learning-to-rank datasets. \textsc{Object} is the
slowest implementation and \textsc{Struct} is slightly faster. 
On the LETOR dataset, \textsc{Struct}
is only slightly slower than \textsc{Struct$^+$}, but on MSLR,
\textsc{Struct$^+$} is faster than \textsc{Struct} by a larger margin in most cases.
\textsc{VPred} outperforms all other techniques, including \textsc{CodeGen}
on MSLR, but is slower than \textsc{CodeGen} on LETOR (except for the
shallowest trees). However, note that in terms of NDCG, Table~\ref{table:effectiveness}(a)
shows no difference in effectiveness, so there is no advantage to
building deeper trees for LETOR.

The conclusion appears clear:\ for tree-based ensembles on
real-world learning-to-rank datasets, we can achieve the best of both
worlds. With a combination of predication and vectorization, we can
make predictions faster than statically-generated if-else blocks, yet retain the
flexibility in being able to specify the model dynamically, which
enables rapid experimentation.

\section{Discussion and Future Work}
\label{section:discussion}

\begin{table}[t]
\caption{Average percentage of examined features (variance in parentheses)
across five folds using various max-number-of-leaves settings.
\label{table:features}}
\begin{center}
\subtable[LETOR-MQ2007]{
\begin{tabular}{|p{1.5cm}|p{0.5cm}|p{0.5cm}|p{0.5cm}|p{0.5cm}|p{0.5cm}|}
\hline
& 3 & 5 & 7 & 9 & 11 \\
\hline
\hline
Percentage of features & 76.7 (5.0) & 72.2 (8.8) & 80.2 (5.6) & 77.6 (7.6) & 84.8 (1.9) \\
\hline
\end{tabular}}
\subtable[MSLR-WEB10K]{
\begin{tabular}{|p{1.5cm}|p{0.6cm}|p{0.6cm}|p{0.6cm}|p{0.6cm}|}
\hline
& 10 & 30 & 50 & 70 \\
\hline
\hline
Percentage of features & 92.7 (1.7) & 96.5 (1.1) & 96.3 (1.9) & 95.6 (1.6) \\
\hline
\end{tabular}}
\end{center}
\vspace{-0.5cm}
\end{table}

Our experiments show that predication and vectorization are effective
techniques for substantially increasing the performance of tree-based models, but
one potential objection might be:\ are we measuring the right thing?
In our experiments, prediction time is measured from when the feature
vector is presented to the model to when the prediction is
made. Critically, we assume that features have already been
computed. What about an alternative architecture where features are
computed lazily, i.e., only when the predicate at a tree node needs to
access a particular feature?

This alternative architecture, where features are computed on demand, is
difficult to study since results will be highly dependent on the
implementation of feature extraction---which in turn
depends on the underlying data structures (layout of the inverted
indexes), compression techniques, and how computation-intensive
the features are. However, there is a much easier
way to study this issue---we can trace the execution of the full tree
ensemble and keep track of the fraction of features that are
accessed. If during the course of making a prediction, most of the
features are accessed, then there is little waste in computing all the
features first and then presenting the complete feature vector to the
model. 

Table~\ref{table:features} shows the average fraction of
features accessed in the final learned models for both
learning-to-rank datasets, with different max leaves
configurations. It is clear that, for both datasets, most of the
features are accessed during the course of making a prediction, and in
the case of the MSLR dataset, nearly all the features are accessed all
the time (especially with deeper trees, which yield higher effectiveness).
Therefore, it makes sense to separate feature extraction
from prediction. In fact, there are independent compelling reasons to do
so:\ a dedicated feature extraction stage can benefit from better reference
locality (when it comes to document vectors, postings, or whatever underlying
data structures are necessary for computing features). 
Interleaving feature extraction with tree traversal may lead
to ``cache churn'', where a particular data structure is repeatedly
loaded and then displaced by other data.

Returning to a point in the introduction:\ 
do these optimizations
actually matter, in the broader context of real-world search
engines? This is of course a difficult question to answer and highly
dependent on the actual search architecture, which is a complex
distributed system spanning hundreds of machines or more.
Here, we venture some
rough estimates. From Figure~\ref{figure:results:actual}(b), the MSLR
dataset, we see that compared to \textsc{CodeGen}, \textsc{VPred}
reduces per-instance prediction time from around 40$\mu{s}$ to around
25$\mu{s}$ (for max leaves setting of 50); this translates into
a 38\% reduction in latency per instance. In a web
search engine, the learning to rank algorithm is applied to a
candidate list of documents that is usually generated by other means
(e.g., scoring with BM25 and a static prior). The exact details are
proprietary, but the published literature does provide
some clues. For example, Cambazoglu et
al.~\cite{Cambazoglu_etal_WSDM2010} (authors from Yahoo!)\ experimented with reranking 200
candidate documents to produce the final ranked list of 20 results
(the first two pages of search results). From these numbers, we can
compute the per-query reranking time to be 8ms using the
\textsc{CodeGen} approach and 5ms with \textsc{VPred}. This
translates into an increase from 125 queries per second to 200
queries per second on a single thread for this phase of the search
pipeline. Alternatively, gains from faster prediction can be leveraged
to rerank more results or take advantage of more features.
This simple estimate suggests that our optimizations can make a noticeable
difference in web search, and given that our techniques are
relatively simple---the predication and vectorization optimizations
definitely seem worthwhile.

During the course of our experiments, we noticed that two assumptions
of our implementations did not appear to be fully valid. First, the
\textsc{Pred} and \textsc{VPred} implementations assume fully-balanced
binary trees (i.e., every node has a left and a right child). In contrast,
recall that \textsc{Struct$^+$} makes no such assumption because with
the left and right pointers we can tightly pack the tree nodes. The
fully-balanced tree assumption does
not turn out to be valid for GBRTs---the learner does not have a
preference for any particular tree topology, and so the trees are unbalanced
most of the time. To compensate for this, the \textsc{Pred} and \textsc{VPred} implementations require
insertion of dummy nodes to
create a fully-balanced tree. Second, we assume that all paths are
equally likely in a tree, i.e., that at each node, the left and right
branches are taken with roughly-equal frequency. We noticed, however,
that this is often not the case. To the extent that one branch is
favored over another, branch prediction provides non-predicated
implementations (i.e., if-else blocks) an advantage, since branch prediction
will guess correctly more often, thus avoiding pipeline flushes.

One promising future direction to address the above two issues is to
adapt the model learning process to prefer balanced trees and
predicates that divide up the feature space evenly. We believe this
can be incorporated into the learning algorithm as a penalty, much in
the same way that regularization is performed on the objective in standard
machine learning. Thus, it is
perhaps possible to jointly learn models that are both fast and good, as
in the recently-proposed
``learning to {\it efficiently} rank'' framework~\cite{Wang_etal_SIGIR2011,XuZhixiang_etal_ICML2012}.

\section{Conclusion}
\label{section:conclusion}

Modern processor architectures are incredibly complex because
technological improvements have been uneven. This paper focuses on one
particular issue:\ not all memory references are equally fast, and in
fact, latency can differ by an order of magnitude. There are a number
of mechanisms to mask these latencies, although it largely depends on
developers knowing how to exploit these mechanisms. The database
community has been exploring these issues for quite some time now, and
in this respect the information retrieval, machine learning, and data
mining communities are behind. 

In this paper, we demonstrate that two relatively simple techniques,
predication and vectorization, along with more efficient memory layouts, can significantly accelerate prediction
performance for tree-based models, both on synthetic data and on
real-world learning-to-rank datasets. Our work explores
architecture-conscious implementations of a particular machine
learning model---but we believe there are plenty of similar
opportunities in other areas of machine learning as well.

\section{Acknowledgments}

This work has been supported by NSF under awards IIS-0916043,
IIS-1144034, and IIS-1218043. Any opinions, findings, conclusions, or
recommendations expressed are the authors' and do not necessarily
reflect those of the sponsor. The first author's deepest gratitude
goes to Katherine, for her invaluable encouragement and wholehearted
support. The second author is grateful to Esther and Kiri for their
loving support and dedicates this work to Joshua and Jacob.

\bibliographystyle{abbrv}

\begin{thebibliography}{10}

\bibitem{Ailamaki_etal_VLDB1999}
A.~Ailamaki, D.~DeWitt, M.~Hill, and D.~Wood.
\newblock {DBMSs} on a modern processor: Where does time go?
\newblock {\em Proceedings of the 25th International Conference on Very Large
  Data Bases (VLDB1999)}, pp. 266--277, Edinburgh, Scotland, 1999.

\bibitem{August_etal_1997}
D.~August, W.~Hwu, and S.~Mahlke.
\newblock A framework for balancing control flow and predication.
\newblock {\em Proceedings of the 30th Annual ACM/IEEE International Symposium
  on Microarchitecture (MICRO1997)}, pp. 92--103, Research Triangle Park, North
  Carolina, 1997.

\bibitem{Boncz_etal_CACM2008}
P.~Boncz, M.~Kersten, and S.~Manegold.
\newblock Breaking the memory wall in {MonetDB}.
\newblock {\em Communications of the ACM}, 51(12):77--85, 2008.

\bibitem{Boncz_etal_CIDR2005}
P.~Boncz, M.~Zukowski, and N.~Nes.
\newblock {MonetDB/X100}: Hyper-pipelining query execution.
\newblock {\em Proceedings of the 2nd Biennial Conference on Innovative Data
  Systems Research (CIDR2005)}, Asilomar, California, 2005.

\bibitem{Burges_2010}
C.~Burges.
\newblock From {RankNet} to {LambdaRank} to {LambdaMART}: An overview.
\newblock Technical Report MSR-TR-2010-82, Microsoft Research, 2010.

\bibitem{Cambazoglu_etal_WSDM2010}
B.~Cambazoglu, H.~Zaragoza, O.~Chapelle, J.~Chen, C.~Liao, Z.~Zheng, and
  J.~Degenhardt.
\newblock Early exit optimizations for additive machine learned ranking
  systems.
\newblock {\em Proceedings of the 3rd ACM International Conference on Web
  Search and Data Mining (WSDM 2010)}, pp. 411--420, New York, 2010.

\bibitem{Criminisi_etal_2011}
A.~Criminisi, J.~Shotton, and E.~Konukoglu.
\newblock Decision forests: A unified framework for classification, regression,
  density estimation, manifold learning and semi-supervised learning.
\newblock {\em Foundations and Trends in Computer Graphics and Vision},
  7(2--3):81--227, 2011.

\bibitem{Ganjisaffar_etal_SIGIR2011}
Y.~Ganjisaffar, R.~Caruana, and C.~Lopes.
\newblock Bagging gradient-boosted trees for high precision, low variance
  ranking models.
\newblock {\em Proceedings of the 34th Annual International ACM SIGIR
  Conference on Research and Development in Information Retrieval (SIGIR2011)},
  pp. 85--94, Beijing, China, 2011.

\bibitem{Ge_Wong_2008}
G.~Ge and G.~Wong.
\newblock Classification of premalignant pancreatic cancer mass-spectrometry
  data using decision tree ensembles.
\newblock {\em BMC Bioinformatics}, 9:275, 2008.

\bibitem{Jacob_2009}
B.~Jacob.
\newblock {\em The Memory System: You Can't Avoid It, You Can't Ignore It, You
  Can't Fake It}.
\newblock Morgan \& Claypool Publishers, 2009.

\bibitem{Jarvelin_Kekalainen_TOIS2002}
K.~{J\"{a}rvelin} and J.~{Kek\"{a}l\"{a}inen}.
\newblock Cumulative gain-based evaluation of {IR} techniques.
\newblock {\em ACM Transactions on Information Systems}, 20(4):422--446, 2002.

\bibitem{Johnson_etal_2012}
N.~Johnson, G.~Zhao, E.~Hunsader, J.~Meng, A.~Ravindar, S.~Carran, and
  B.~Tivnan.
\newblock Financial black swans driven by ultrafast machine ecology.
\newblock {\em arXiv:1202.1448v1}, 2012.

\bibitem{KimHyesoon_etal_2006}
H.~Kim, O.~Mutlu, Y.~Patt, and J.~Stark.
\newblock Wish branches: Enabling adaptive and aggressive predicated execution.
\newblock {\em IEEE Micro}, 26(1):48--58, 2006.

\bibitem{LiHang_2011}
H.~Li.
\newblock {\em Learning to Rank for Information Retrieval and Natural Language
  Processing}.
\newblock Morgan \& Claypool Publishers, 2011.

\bibitem{Olukotun_Hammond_2005}
K.~Olukotun and L.~Hammond.
\newblock The future of microprocessors.
\newblock {\em ACM Queue}, 3(7):27--34, 2005.

\bibitem{Panda_etal_VLDB2009}
B.~Panda, J.~Herbach, S.~Basu, and R.~Bayardo.
\newblock {PLANET}: Massively parallel learning of tree ensembles with
  {MapReduce}.
\newblock {\em Proceedings of the 35th International Conference on Very Large
  Data Bases (VLDB2009)}, pp. 1426--1437, Lyon, France, 2009.

\bibitem{RaoJun_Ross_VLDB1999}
J.~Rao and K.~Ross.
\newblock Cache conscious indexing for decision-support in main memory.
\newblock {\em Proceedings of the 25th International Conference on Very Large
  Data Bases (VLDB1999)}, pp. 78--89, Edinburgh, Scotland, 1999.

\bibitem{RossKenneth_etal_2005}
K.~Ross, J.~Cieslewicz, J.~Rao, and J.~Zhou.
\newblock Architecture sensitive database design: Examples from the {Columbia}
  group.
\newblock {\em Bulletin of the Technical Committee on Data Engineering},
  28(2):5--10, 2005.

\bibitem{Schietgat_etal_2010}
L.~Schietgat, C.~Vens, J.~Struyf, H.~Blockeel, D.~Kocev, and S.~{D\u{z}eroski}.
\newblock Predicting gene function using hierarchical multi-label decision tree
  ensembles.
\newblock {\em BMC Bioinformatics}, 11:2, 2010.

\bibitem{Svore_Burges_2011}
K.~Svore and C.~Burges.
\newblock Large-scale learning to rank using boosted decision trees.
\newblock {\em Scaling Up Machine Learning}. Cambridge University Press, 2011.

\bibitem{Tyree_etal_WWW2011}
S.~Tyree, K.~Weinberger, and K.~Agrawal.
\newblock Parallel boosted regression trees for web search ranking.
\newblock {\em Proceedings of the 20th International Conference on World Wide
  Web (WWW2011)}, pp. 387--396, Hyderabad, India, 2011.

\bibitem{Wang_etal_SIGIR2011}
L.~Wang, J.~Lin, and D.~Metzler.
\newblock A cascade ranking model for efficient ranked retrieval.
\newblock {\em Proceedings of the 34th Annual International ACM SIGIR
  Conference on Research and Development in Information Retrieval (SIGIR2011)},
  pp. 105--114, Beijing, China, 2011.

\bibitem{XuZhixiang_etal_ICML2012}
Z.~Xu, K.~Weinberger, and O.~Chapelle.
\newblock The greedy miser: Learning under test-time budgets.
\newblock {\em Proceedings of the 29th International Conference on Machine
  Learning (ICML 2012)}, Edinburgh, Scotland, 2012.

\bibitem{Zukowski_etal_2005}
M.~Zukowski, P.~Boncz, N.~Nes, and S.~{H\'{e}man}.
\newblock {MonetDB/X100}---a {DBMS} in the {CPU} cache.
\newblock {\em Bulletin of the Technical Committee on Data Engineering},
  28(2):17--22, 2005.

\end{thebibliography}

\end{document}